\documentclass[10pt]{article}

\usepackage{fullpage}
\usepackage{setspace}
\usepackage{parskip}
\usepackage{titlesec}
\usepackage{xcolor}
\usepackage{lineno}
\usepackage{authblk}
\usepackage{graphicx}
\usepackage[space]{grffile}
\usepackage{latexsym}
\usepackage{textcomp}
\usepackage{longtable}
\usepackage{tabulary}
\usepackage{booktabs,array,multirow}
\usepackage{amsfonts,amsmath,amssymb,amsthm}
\usepackage{bm}%
\usepackage{siunitx}
\usepackage{color}
\usepackage{mathtools}
\usepackage[normalem]{ulem} 
\usepackage[sc,osf]{mathpazo}

\PassOptionsToPackage{hyphens}{url}
\usepackage[]{hyperref}
\usepackage{etoolbox}
\makeatletter
\patchcmd\@combinedblfloats{\box\@outputbox}{\unvbox\@outputbox}{}{%
	\errmessage{\noexpand\@combinedblfloats could not be patched}%
}%
\makeatother

\renewenvironment{abstract}
{{\bfseries\noindent{\abstractname}\par\nobreak}\footnotesize}
{\bigskip}

\titlespacing{\section}{0pt}{*3}{*1}
\titlespacing{\subsection}{0pt}{*2}{*0.5}
\titlespacing{\subsubsection}{0pt}{*1.5}{0pt}

\newif\iflatexml\latexmlfalse

\AtBeginDocument{\DeclareGraphicsExtensions{.pdf,.PDF,.eps,.EPS,.png,.PNG,.tif,.TIF,.jpg,.JPG,.jpeg,.JPEG}}

\usepackage[T1]{fontenc}
\usepackage[english]{babel}

\graphicspath{{figure/}}

\begin{document}
	
	\title{The onset of dehydrogenation in solid Ammonia Borane, an \textit{ab-initio} metadynamics study}
	
	\author[1,*]{Valerio Rizzi}%
	\author[1,**]{Daniela Polino}
	\author[2,***]{Emilia Sicilia}
	\author[2]{Nino Russo}
	\author[1,****]{Michele Parrinello}
	\affil[1]{Department of Chemistry and Applied Biosciences, ETH Zurich, and Facolt\`a di Informatica, Istituto di Scienze Computazionali, Universit\`a della Svizzera Italiana, Via G. Buffi 13, 6900 Lugano, Switzerland}%
	\affil[2]{Dipartimento di Chimica e Tecnologie Chimiche, Universit\`a della Calabria, 87036 Rende (CS), Italy}
	\affil[*]{valerio.rizzi@usi.ch}
	\affil[**]{daniela.polino@usi.ch}
	\affil[***]{emilia.sicilia@unical.it}
	\affil[****]{michele.parrinello@phys.chem.ethz.ch}
	
	\vspace{-1em}
	
	\date{\today}
	
	\begingroup
	\let\center\flushleft
	\let\endcenter\endflushleft
	\maketitle
	\endgroup
	
	\selectlanguage{english}

\begin{abstract}

The discovery of effective hydrogen storage materials is fundamental for the progress of a clean energy economy. 
Ammonia borane ($\mathrm{H_3BNH_3}$) has attracted great interest as a promising candidate but the reaction path that leads from its solid phase to hydrogen release is not yet fully understood. 
To address the need for insights in the atomistic details of such a complex solid state process, in this work we use \textit{ab-initio} molecular dynamics and metadynamics to study the early stages of AB dehydrogenation.
We show that the formation of ammonia diborane ($\mathrm{H_3NBH_2(}$$\mu$$\mathrm{-H)BH_3}$) leads to the release of $\mathrm{NH_4^+}$, which in turn triggers an autocatalytic $\mathrm{H_2}$ production cycle. 
Our calculations provide a model for how complex solid state reactions can be theoretically investigated and rely upon the presence of multiple ammonia borane molecules, as substantiated by standard quantum-mechanical simulations on a cluster.

\end{abstract}

\begin{figure}[tb]
	\begin{center}
		\includegraphics[width=1.0\columnwidth]{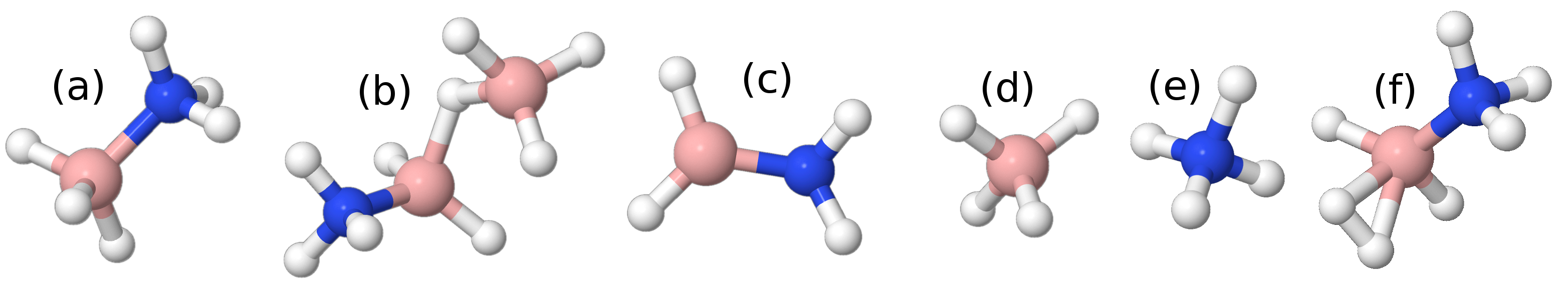}
		\caption{\label{fig:molecole}
			Structure of (a) ammonia borane (AB) $\mathrm{H_3BNH_3}$, (b) ammonia diborane (AaDB) $\mathrm{H_3NBH_2(}$$\mu$$\mathrm{-H)BH_3}$, (c) monomeric aminoborane (AB2) $\mathrm{H_2BNH_2}$, (d) borohydride anion $\mathrm{BH_4^-}$, (e) ammonium cation $\mathrm{NH_4^+}$ and (f) ammoniaborocation ($\mathrm{ABH^+}$) $\mathrm{H_4^+BNH_3}$, where the identity of the hydrogen atoms forming the $\mathrm{H_2}$ moiety is interchangeable.
		}
	\end{center}	
\end{figure}

One of the major obstacles to a wider application of hydrogen as energy vector is the development of safe and efficient storage media.\cite{Huang2012,Pedicini2018} For this reason, chemical storage is considered to be a viable alternative and, amongst chemical hydrides, one of the most studied candidates continues to be ammonia borane ($\mathrm{H_3BNH_3}$, AB) Fig. \ref{fig:molecole} (a).

The AB molecule presents a high gravimetric hydrogen density and is isoelectronic to ethane $\mathrm{H_3C-CH_3}$, with the crucial difference of having a weaker dative boron-nitrogen bond whose energy is less than one third of ethane's carbon-carbon bond strength.\cite{Grant2009a,Weismiller2010b,Demirci2017} The polar nature of the B-N bond makes the N-H hydrogen atoms protic and the B-H hydrogen atoms hydridic. It forms a stable molecular solid at room temperature and 
its polarity leads to the formation of strong intermolecular dihydrogen bonds  
N-$\mathrm{H}^{\delta^+}\cdots\mathrm{\prescript{\delta^-}{}{H}}$-B \cite{Morrison2004} that significantly influence the properties of such a compound.

When close to its melting point of $T_\mathrm{m} \sim 370$K, solid AB goes to an intermediate enhanced mobility state AB* \cite{Stowe2007,Shaw2010} and, after an induction time that depends on preparation, $\mathrm{H_2}$ molecules are released. \cite{Huang2012,Hess2009,Staubitz2010} The system can generate as much as one $\mathrm{H_2}$ per formula unit and what is left is a disordered mixture of polymeric $(\mathrm{-H_2BNH_2-})_n$ aminoboranes (PAB) that further decompose.\cite{Demirci2017}
In spite of intense experimental efforts,\cite{Al-Kukhun2013,Babenko2017,Petit2017} there is no general consensus on how dehydrogenation proceeds. 
In the early stages of dehydrogenation, the formation of diammoniate of diborane ($\mathrm{[(NH_3)_2BH_2^+][BH_4^-]}$, DADB) \cite{Stowe2007} and ammonia diborane ($\mathrm{H_3NBH_2(}$$\mu$$\mathrm{-H)BH_3}$, AaDB) Fig. \ref{fig:molecole} (b) \cite{Chen2011} has been reported.
Another relevant experimental information is the fact that, upon doping, dehydrogenation is dramatically enahnced.\cite{Heldebrant2008}

Standard gas phase quantum-chemical calculations have been performed on what have been hypothesized to be the initial dehydrogenation steps.\cite{Nguyen2007a,Zimmerman2009,Vijayalakshmi2017} Theoretical and experimental findings support the hypothesis that, in the first step, one of the B-N bonds must be broken, leading to dimeric species formation.\cite{Nguyen2007,Al-Kukhun2013,Babenko2017} The details of the mechanism by which molecular hydrogen is further released, however, are still unclear.\cite{Staubitz2010,Demirci2017}

This somewhat uncertain scenario has prompted us to simulate a periodically repeated system at temperatures close to $T_\mathrm{m}$. Simulations of crystals give, indeed, a more realistic description of the environment in which the actual reaction takes place. Furthermore, they fully take into account the very large anharmonic effects and the changes that take place in the environment as the reaction proceeds. These effects proved to be crucial to obtain a clear picture of the steps leading to the initial $\mathrm{H}_2$ release. 

Performing straightforward \textit{ab-initio} molecular dynamics (AIMD) however would not be sufficient, as the high kinetic barriers that lead to dehydrogenation would significantly limit the effectiveness of such an approach. Metadynamics \cite{Laio2002,Barducci2008,Dama2014,Valsson2016} offers a crucial help in overcoming this limitation, so we have carried out metadynamics simulations to explore the reaction landscape.
Throughout such exploration, we have discovered a bewildering variety of possible reaction channels taking place at different stages of the dehydrogenation process. From all this wealth of data, we have extracted a picture in agreement with experiments \cite{Stowe2007,Heldebrant2008,Demirci2017} involving, after the initial formation of the dimeric species AaDB, an intermediate state where both $\mathrm{NH_4^+}$ and $\mathrm{BH_4^-}$ are present. This very unstable state, that has a lifetime on the millisecond order of magnitude, can decay via a significant number of channels that lead to dehydrogenation. Some of the channels are autocatalytic.

The next dominant steps start with a proton transfer from the $\mathrm{NH_4^+}$ cation to AB. This causes the formation of a fluxional compound that we call ammoniaborocation ($\mathrm{H_4^+BNH_3}$, $\mathrm{ABH^+}$) Fig. \ref{fig:molecole} (f), in analogy with the famous $\mathrm{CH_5^+}$ carbocation \cite{Marx1995} in which protons continuously scramble. A key event paving the way to many of the crucial steps of the reaction is the formation of three-centre two-electron bonds.\cite{DeKock1988} The subsequent dynamics of $\mathrm{ABH^+}$ strongly depends on the environment and can generate a number of products along with the release of $\mathrm{H}_2$.

All the key pathways identified by the metadynamics simulations have been reproduced by using standard quantum-mechanical calculations. It is highly rewarding that some of the transition states (TSs) could be recovered, provided that a sufficiently large number of molecules is taken into account to mimic the environment. Such a study could not have been possible without the insights coming from the \textit{ab-initio} metadynamics simulations.


The presence of a precursor phase AB* and the dehydrogenation acceleration due to doping point to a bulk phenomenon. Thus we used periodic boundary conditions and started the simulations from a perfect crystal arrangement \cite{Hess2009} of 16 AB molecules. We ran AIMD on the code CP2K \cite{VandeVondele2005} and used the PBE functional.\cite{Perdew1996} The system was kept at a temperature of $400$ K via a velocity rescaling thermostat.\cite{Bussi2007} 
This temperature corresponds to the one where dehydrogenation is experimentally observed.
At such a temperature, the timescale required for crossing the dehydrogenation energy barrier is inaccessible to molecular dynamics simulations.

To overcome this limitation, we used metadynamics \cite{Laio2002,Barducci2008,Dama2014,Valsson2016} in its Well-Tempered (WT) version as implemented in the PLUMED plugin.\cite{Tribello2014} In such a way, sampling of the configuration space is enhanced so that the occurrence of chemical reactions is facilitated and takes place in an affordable computational time. In WT metadynamics this is achieved by applying on the fly a bias potential that depends on a selected number of collective variables (CVs), the choice of which is of great importance.

The CVs are functions of the atomic coordinates and are meant to describe the system's slow degrees of freedom whose fluctuations eventually lead to the reactive process. In the choice of CVs, we did not prejudge the outcome of the reaction or the identity of the atoms involved, so that reactions are free to take place on any equivalent site of the system. A detailed description of the CVs used to simulate the various reaction steps can be found in the Supporting information (SI).
In complex situations, we combined a large number of CVs into a smaller one using the recently developed HLDA method.\cite{Mendels2018,Piccini2018} 
In other cases where the system is unlikely to return to the reactant basin, we calculated the rates of the forward reaction using infrequent metadynamics.\cite{Tiwary2013,Salvalaglio2014}


\begin{figure}[tb]
	\begin{center}
		\includegraphics[width=0.8\columnwidth]{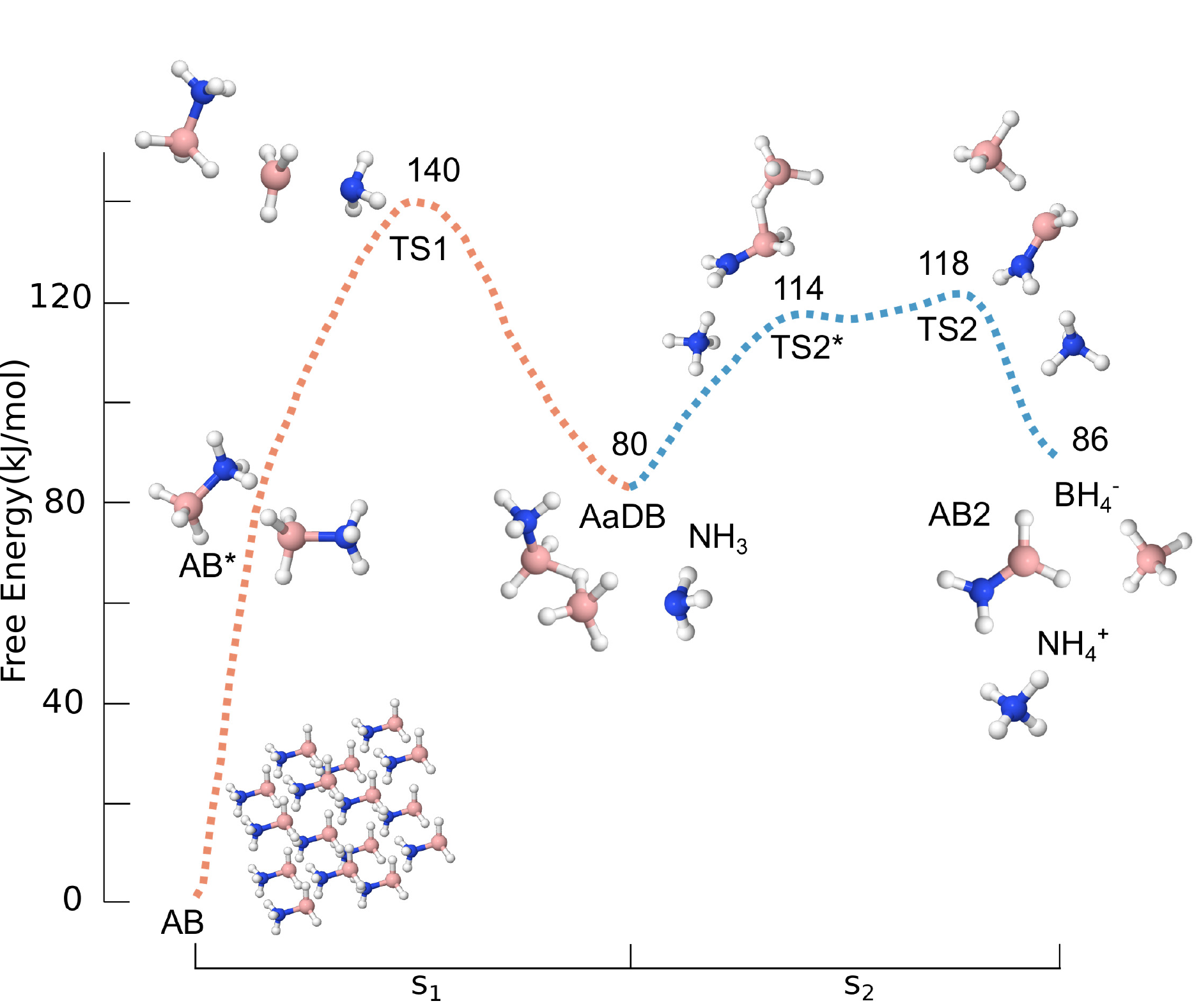}
		\caption{\label{fig:diag}
			Energy level diagram representing the initial sequence of reactions in the pathway leading solid state AB to dehydrogenation. The lowest FE path is juxtaposed to the diagram. The variables $s_1$ and $s_2$ are complex adimensional collective that are used at different stages of the calculation. Further details can be found in the SI.
			The energy values are in kJ/mol with a resolution of 2 kJ/mol.
		}
	\end{center}	
\end{figure}

In our search for a dehydrogenation pathway, we have observed a number of reactions involving different reactants and products. These are described in the SI, while here we report only the steps that lead the solid to an autocatalytic step that we believe to be the main avenue to $\mathrm{H_2}$ production. We started the simulations by heating the system to $T = 400$ K and then introduced a bias to facilitate the first B-N bond cleavage. 
The free energy surface (FES) is reported in Fig. (S.1) as a function of the two driving CVs representing the B-N bond stretch and the formation of a hydrogen bridge between boron atoms.

The corresponding lowest free energy (FE) path is shown in Fig. (\ref{fig:diag}). For clarity, in the diagram only some representative molecules are depicted, without explicitly showing the surrounding AB molecules. The reciprocal molecular positions shown in the TSs are accurate and correspond to configurations observed in the simulations, while the molecular positions of the minima are just illustrative.

Along the reaction path, in the initial step a couple of molecules change their relative orientation such that two $\mathrm{BH_3}$ groups face each other (AB*), indicating that the bulk steric effects do not hinder the reaction. 
As a consequence of this change of orientation, also seen in experiments,\cite{Stowe2007,Shaw2010} a number of stabilizing dihydrogen bonds are broken, allowing the reaction 
\begin{equation}
\mathrm{AB} + \mathrm{AB} \ \longrightarrow \ \mathrm{AaDB} + \mathrm{NH_3}
\end{equation}
to occur. The B-N bond breaking is facilitated by the formation of the AaDB key intermediate and of a $\mathrm{NH_3}$ moiety.
This is in agreement with experiments that have detected the presence of AaDB in the earlier dehydrogenation phases.\cite{Chen2011}
The barrier height for this TS, that is the highest in our scheme, is in line with kinetic experiments.\cite{Demirci2017}

The reaction proceeds according to the scheme
\begin{equation}
\mathrm{AaDB} + \mathrm{NH_3} \ \longrightarrow \ \mathrm{AB2} + \mathrm{NH_4^+} + \mathrm{BH_4^-} 
\end{equation}
overcoming lower energy barriers and leading to the formation of
monomeric aminoborane ($\mathrm{H_2BNH_2}$, AB2) Fig. \ref{fig:molecole} (c), a compound prone to oligomerization.\cite{Zimmerman2009} 
For this step to occur, it is required that the $\mathrm{NH_3}$ molecule comes close to the AaDB nitrogen to allow a proton transfer from AaDB to $\mathrm{NH_3}$ and the cleavage of the hydrogen bridge between the boron atoms to, ultimately, form AB2, $\mathrm{BH_4^-}$ and $\mathrm{NH_4^+}$.

\begin{figure}[tb]
	\begin{center}
		\includegraphics[width=0.6\columnwidth]{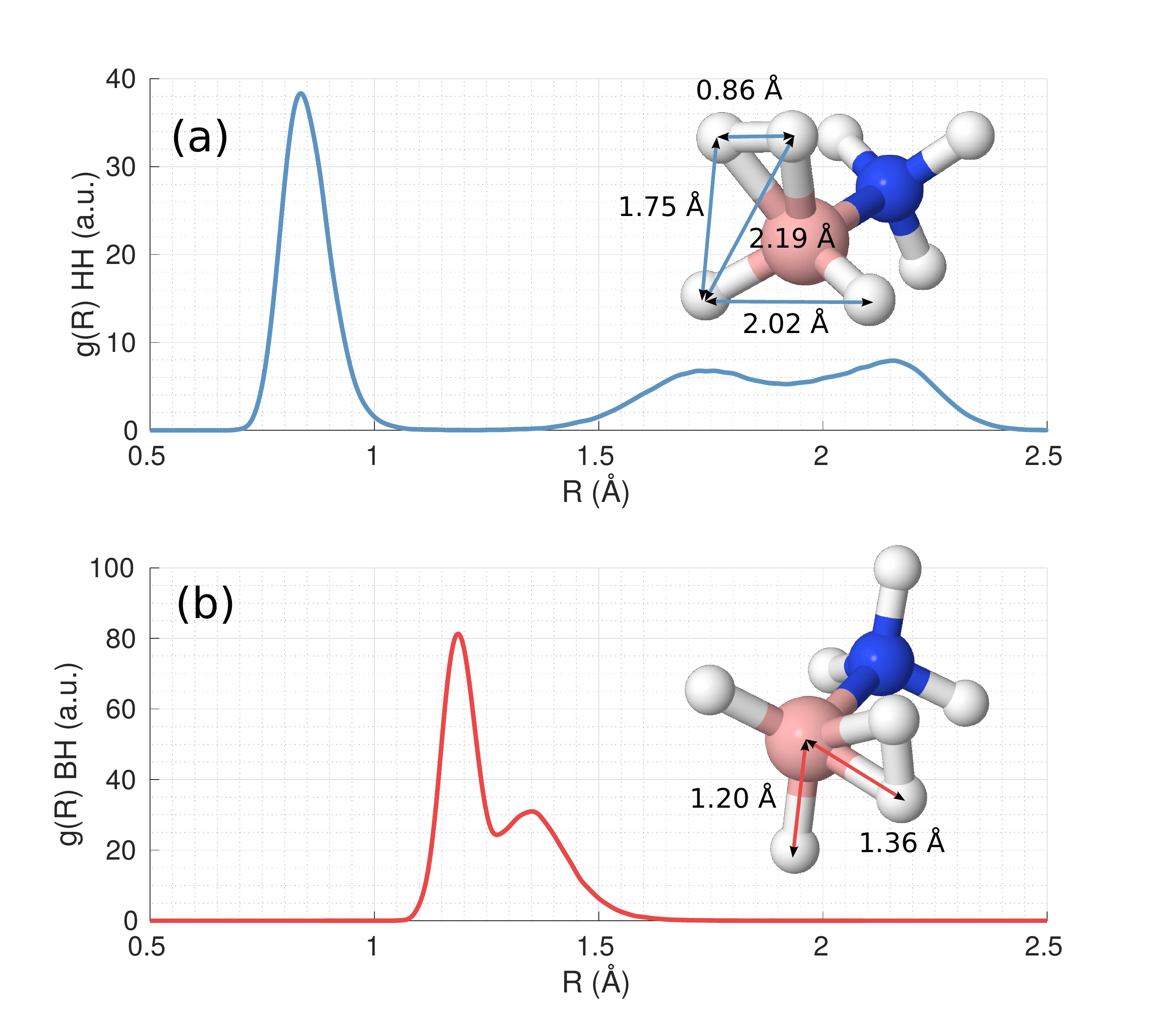}
		\caption{\label{fig:ABH}
		Radial distribution function of (a) the $\mathrm{H}$-$\mathrm{H}$ distances between the $\mathrm{H}$ atoms bound to boron in $\mathrm{ABH^+}$ and (b) the $\mathrm{B}$-$\mathrm{H}$ bond lengths in $\mathrm{ABH^+}$. On the right hand side, the distances corresponding to the peaks are highlighted.
		}
	\end{center}	
\end{figure}
\begin{figure}[tb]
	\begin{center}
		\includegraphics[width=0.9\columnwidth]{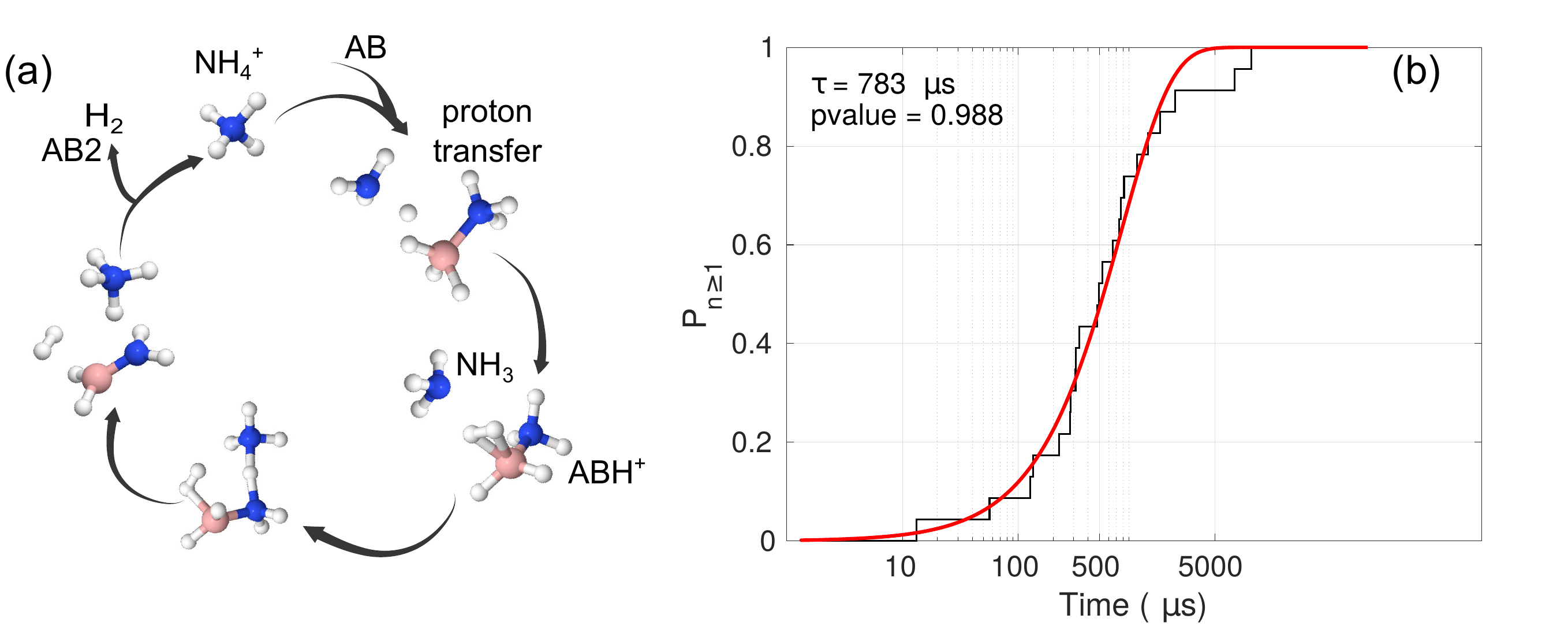}
		\caption{\label{fig:infreq}
			In (a) a sketch of an autocatalytic $\mathrm{H_2}$ release cycle involving the $\mathrm{NH_4^+}$ ion, in (b) the dehydrogenation probability distribution with a Poisson process fit in red.
		}
	\end{center}	
\end{figure}

From this crucial state, a number of different pathways can be originated.
Initially, they all involve the donation of a proton from the ammonium cation to an AB boron, with the formation of the corresponding cation that we name ammoniaborocation ($\mathrm{ABH^+}$). 
One of the most prominent features of this step is once again the appearance of a three-centre two-electron bond.

The structure of $\mathrm{ABH^+}$ is better described statistically, as the identity of the hydrogen atoms participating in the $\mathrm{H_2}$-like bond changes continuously. Thus, we look at the distribution of H-H and B-H distances for the $\mathrm{H}$ atoms bound to the boron in $\mathrm{ABH^+}$. The H-H radial distribution function in Fig. (\ref{fig:ABH}) (a) clearly displays a short distance peak corresponding to the fluxional $\mathrm{H_2}$, while the other hydrogens contribute to the longer distance peaks. Also the B-H bond lengths vary, with the ones relative to $\mathrm{H}_2$ being longer, as shown in Fig. (\ref{fig:ABH}) (b). 

The fate of the unstable $\mathrm{ABH^+}$ depends on the environment, as a large number of alternative reaction channels is viable. Eventually, all routes lead to the release of $\mathrm{H}_2$ with the production of a number of possible compounds, among which AB2, $\mathrm{NH_4^+}$, DADB, AaDB or $\mathrm{BH_5}$. The latter has a fluxional nature analogous to the one of the ammoniaborocation.\cite{Konczol2014} Among the many paths, the autocatalytic process shown in Fig. (\ref{fig:infreq}) (a)
\begin{equation}
\mathrm{NH_4^+} + \mathrm{AB} \ \longrightarrow \ \mathrm{NH_3} + \mathrm{ABH^+} \ \longrightarrow \ \mathrm{NH_4^+} + \mathrm{H_2} + \mathrm{AB2},
\end{equation}
clearly plays a dominant role.

In this work, rather than studying each route individually, we have thought it more useful to calculate the lifetime of the state where $\mathrm{NH_4^+}$ and $\mathrm{BH_4^-}$ are present. The notion of state here should be taken in a rather broader sense than normally in quantum chemistry. Indeed, the $\mathrm{NH_4^+}$ and $\mathrm{BH_4^-}$ ions can freely diffuse, as their position in the solid is not fixed and the rest of the ammonia borane crystal is an integral part of the state definition.

By employing infrequent metadynamics,\cite{Tiwary2013,Salvalaglio2014} we estimated $\tau \sim 0.8$ ms, as shown in the distribution of dehydrogenation events in Fig. (\ref{fig:infreq}) (b). 
The presence of multiple dehydrogenation routes can explain the complex composition of the residue experimentally detected.\cite{Demirci2017} A list of the observed reactions is given in the SI.

It is remarkable that, even though dehydrogenation could in principle start from any molecule in the system, all the observed events originate from the action of the ammonium ion. Indeed, by manually excluding $\mathrm{NH_4^+}$ from the reaction environment, analogous dehydrogenation mechanisms involving $\mathrm{BH_4^-}$ could be observed, albeit with significantly higher energy barriers.

\begin{figure}[tb]
	\begin{center}
		\includegraphics[width=1.0\columnwidth]{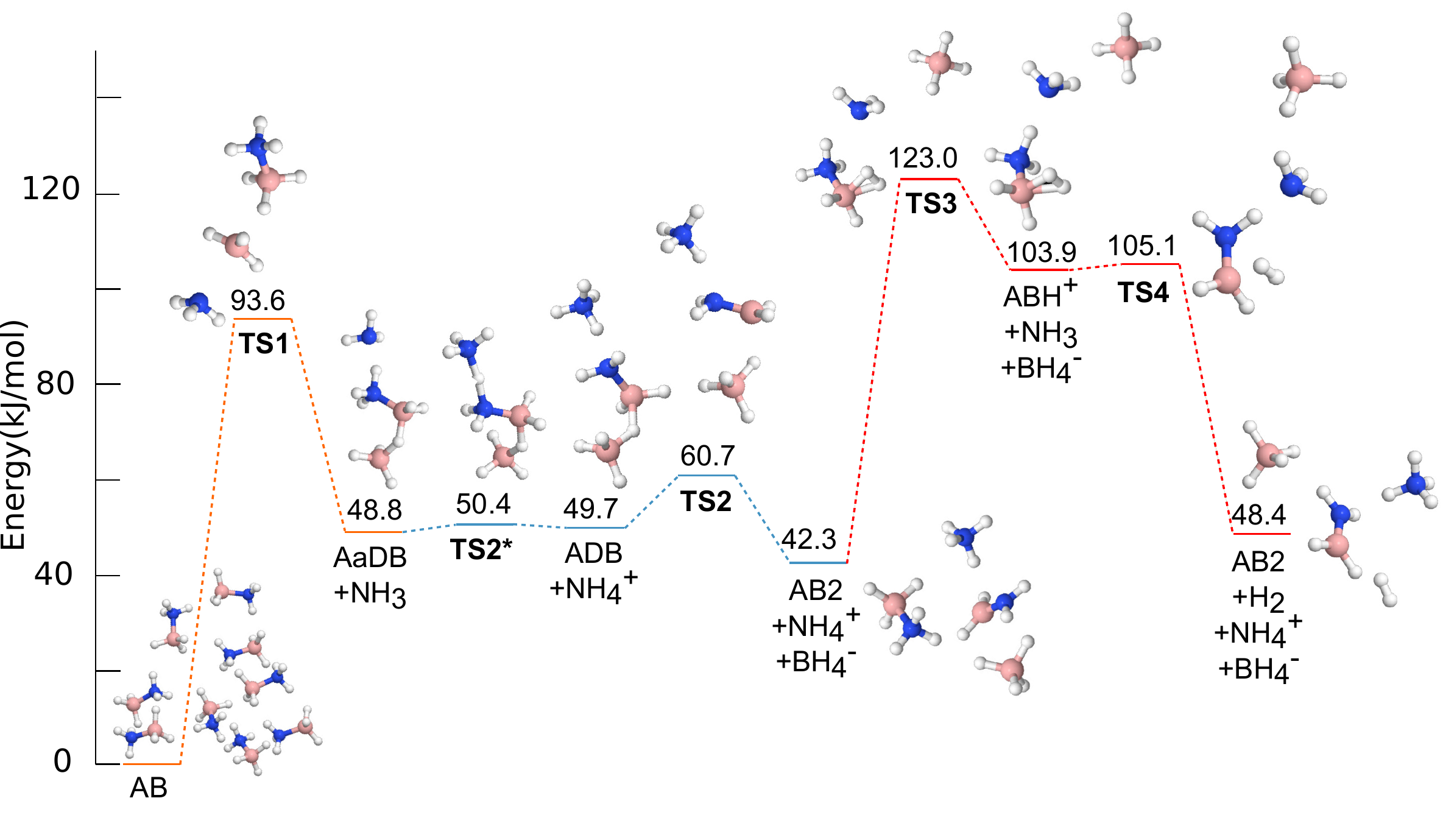}		
		\caption{\label{fig:static}
			B3LYP-D3 free energy profile for a cluster of 9 AB molecules illustrating the reaction steps leading to (orange line) the formation of the first AaDB intermediate and the release of the $\mathrm{NH_4^+}$ and $\mathrm{BH_4^-}$ ions together with an aminoborane molecule AB2 and (blue line) the further reaction of the ammonium ion with an AB molecule to release molecular hydrogen (red line).
			The relative free energies are expressed in kJ/mol and have been calculated with respect to the free energy of the initially interacting 9 molecules. Fully optimized structures of minima and TSs located along the reported FE profiles are sketched.
		}
	\end{center}	
\end{figure}

In order to validate the AIMD findings and assess the effect of using an upgraded exchange and correlation functional, we have performed standard static quantum-mechanical simulations. 
Such calculations have been carried out using the Gaussian 09 suite of programs, employing the hybrid Becke three-parameter exchange functional \cite{Becke1993} and the Lee-Yang-Parr correlation functional \cite{Lee1988} B3LYP, including dispersion corrections for non-bonding interaction through the Grimme approach.\cite{Grimme2010} Standard 6-311+G** Pople basis sets have been used to describe all the atoms. A cluster of 9 AB molecules has been used to simulate the complex environment of a melting solid. More details can be found in the SI.

These simulations proved to be capable of locating the minima and TSs described earlier, provided that a sufficiently large number of molecules was explicitly included to mimic the extended environment. 
Our calculations have been performed on a cluster of 9 AB molecules and the resulting free energy profile is shown in Fig. (\ref{fig:static}). Only the first TS (\textbf{TS1}) has been previously intercepted by Nguyen et al. \cite{Nguyen2007a} in simulations involving two AB molecules, whereas for all the other TSs a larger number of molecules has been required.

Focusing on two interacting AB molecules, \textbf{TS1} involves the attack of the atom in one AB to the H atom of the $\mathrm{BH_3}$ in another AB, leading to the formation of a B-H-B bond bridge and causing the breaking of the B-N bond in the attacking molecule. Overcoming the corresponding barrier of 93.6 kJ/mol, the species AaDB is formed. The reaction proceeds with the formation of two $\mathrm{NH_4^+}$ and $\mathrm{BH_4^-}$ ions together with $\mathrm{H_2 BN H_2}$ in a stepwise manner going through two TSs (\textbf{TS2}* and \textbf{TS2}) and a very shallow minimum corresponding to the transfer of a proton from the AaDB $\mathrm{NH_4^+}$ group to the free ammonia molecule.

The subsequent steps, selected as the most significant on the basis of metadynamics simulations, involve the transfer of a proton from the $\mathrm{NH_4^+}$ ion to the $\mathrm{BH_3}$ group of an AB molecule to yield the key fluxional $\mathrm{ABH^+}$ species, overcoming an activation energy barrier of 83.2 kJ/mol. From $\mathrm{ABH^+}$, immediately a protic N-H hydrogen is transferred to the formed ammonia molecule, causing the simultaneous detachment of the pre-formed $\mathrm{H_2}$ molecule.



Our findings provide an answer to the long-standing problem of how solid state AB releases $\mathrm{H_2}$ close to melting and also represent a methodological scheme for uncovering the most viable reaction channels involving all relevant intermediates and elementary steps needed for the accurate description of chemical processes occurring in a complex environment.
Through metadynamics simulations, we identified intermediate compounds such as AaDB and $\mathrm{N H_4^+}$ and provided an estimate of the energy barriers underlying their formation. The cleavage of one of the B-N bonds and the formation of the key AaDB species present an activation barrier of about 140 kJ/mol and represent the rate limiting event, while the ensuing formation of $\mathrm{N H_4^+}$ requires a lower barrier of about 40 kJ/mol. From that state, whose mean lifetime we estimated to be in the millisecond range, dehydrogenation can proceed through a number of channels. 
All of these channels start from $\mathrm{N H_4^+}$ leading to the formation of the fluxional compound $\mathrm{ABH^+}$ and some of them are autocatalytic.
The complex nature of the dehydrogenation reaction can explain the large product distribution that is experimentally detected. 
An analogous path could be described in terms of minima and TSs through static calculations, provided that enough AB molecules are taken into consideration to mimic an extended environment. 



\selectlanguage{english}

\clearpage

\bibliography{PAPERS_USI_2017-Ammonia_Borane,PAPERS_USI_2017-Metadinamica,../extra}{}
\bibliographystyle{naturemag.bst} 

\section*{Acknowledgements}
We acknowledge the Swiss National Science Foundation Grant Nr. 200021\textunderscore169429/1 and the European Union Grant Nr. ERC-2014-AdG-670227/VARMET for funding.
The authors also thank the Universit\`a della Calabria for financial support. 
The dynamic simulations were carried out on the ETH Euler cluster and on the M{\"{o}nch cluster at the Swiss National Super-computing Center (CSCS).

\end{document}